# Valid p-Values and Expectations of p-Values Revisited

Albert Vexler

*Department of Biostatistics, The State University of New York at Buffalo, Buffalo, NY 14214, U.S.A*

E-mail: *avexler@buffalo.edu*

**Abstract** A storm of favorable or critical publications regarding p-values-based procedures has been observed in both the theoretical and applied literature. We focus on valid definitions of p-values in the scenarios when composite null models are in effect. A valid p-value (VpV) statistic can be used to make a prefixed level-$\alpha$ decision. In this context, Kolmogorov Smirnov goodness-of-fit tests and the normal two sample problem are considered. In particular, we examine an issue regarding the goodness-of-fit testability based on a single observation. This article exemplifies constructions of new test procedures, advocating practical reasons to implement VpV-based mechanisms. The VpV framework induces an extension of the conventional expected p-value (EPV) tool for measuring the performance of a test. Associating the EPV concept with the receiver operating characteristic (ROC) curve methodology, a well-established biostatistical approach, we propose a Youden's index based optimality principle to derive critical values of decision making procedures. In these terms, the significance level $\alpha = 0.05$ can be suggested, in many situations. In light of an ROC curve analysis, we introduce partial EPV's to characterize properties of tests including their unbiasedness. We also provide the intrinsic relationship between the Bayes Factor (BF) test statistic and the BF of test statistics.

## 1. Introduction

Commonly, statistical testing procedures are designed to draw a conclusion (or make an action) with respect to the binary decision of rejecting or not rejecting the null hypothesis $H_0$, depending on locations of the corresponding values of the observed test statistics, i.e. detecting whether test statistics' values belong to a fixed sphere or interval. Oftentimes, p-values can serve as a data-driven approach for testing statistical hypotheses based on using the observed values of test statistics as the thresholds in the theoretical probability of the Type I error. P-values can themselves also serve as a summary type result based on data in that they provide meaningful data based evidence about the null hypothesis. This principle simplifies and standardizes statistical decision-making policies. In this manner, for example, different algorithms for combining decision-making rules using their p-values as test statistics can be naturally derived.

Data driven research oriented journals have started to alarm regarding critical issues that have occurred in experimental studies where statistical decision-making procedures have been involved and p-values-based conclusions have been misused and/or misinterpreted. This has stimulated a storm of favorable or critical publications regarding p-values-based procedures in both the theoretical and applied literature (e.g., Wasserstein and Lazar, 2016; Ionides et al., 2017).

In this article we indicate that proper uses of p-values depend on their structures that can be built in different manners. Our aim is to describe the following specific areas: (1) Valid definitions of p-values in parametric and nonparametric settings; (2) Examples of new VpV–based test procedures that are reasonable to be implemented in practice; (3) Examination of the issue to test for goodness-of-fit based on a single observation (the corresponding motivations are presented below); (4) Revisiting the EPV concept in order to attend to VpV's and optimal selections of tests' significance levels; (5) Proposing partial EPV's to characterize properties of decision making mechanisms; and (6) Demonstrating an interesting fact that the BF based on the BF test statistic comes to be the BF.

In particular, we concentrate on tests in the presence of nuisance parameters. In parametric statistical statements, Berger and Boos (1994) and then Silvapulle (1996) studied testing problems related to a model with an unknown parameter, say $\theta$, under $H_0$. In the frequentist hypothesis testing fashion a test is deemed statistically significant if the p-value is below some threshold known as the significance level, say $\alpha \in [0,1]$. In this context, if $\theta$ were known, then the Type I error rate depends on values of $\theta$, and we can define the conventional p-value such that, under the null hypothesis, $\Pr(p-value \leq \alpha / H_0) \leq \alpha$, for each $\alpha \in [0,1]$. When the value of $\theta$ is unspecified, $H_0$ is no longer simple, however we aim to construct a p-value, preserving the property $\Pr(p-value \leq \alpha / H_0) \leq \alpha$ that provides the corresponding test to be a level-$\alpha$ test. In this case, a statistic 'p-value' is called a valid p-value (VpV). Note that, if the VpV property is not deemed desirable, there are different ways to handle the unknown $\theta$, e.g., estimating values of $\theta$ or using posterior predictive p-values (Bayarri and Berger, 2000). We investigate the concept of VpV's with respect to nonparametric classical Kolmogorov Smirnov (KS) tests for goodness-of-fit and the normal two sample problem. In this framework, we consider the problem of constructing level-$\alpha$ goodness-of-fit tests based on KS measures of the discrepancy between a single observed data point and the value expected under the null model. This issue is not trivial and requires attention from both theoretical and experimental points of view (see Section 2.1 for details). In this context, e.g., we can make mention of an outstanding study of Portnoy (2019) related to problems of making statistical inference based on a single observation.

For example, in practice, the high cost associated with measuring biomarkers values can significantly restrict further biostatistical applications. When analysis is restricted by the high cost of assays, one can suggest applying an efficient pooling design for collection of data (see, e.g., Vexler et al., 2008; Schisterman, Vexler, 2008 and Schisterman et al., 2011, for details). Pooled data can also be an organic output of a study. In order to reduce the cost or labor intensiveness of a study, a pooling strategy may be employed whereby $m \geq 2$ individual specimens are physically combined into a single 'pooled' unit for analysis. Thus, applying pooling design provides a $m$-fold decrease in the number of measured assays. Each pooled sample test result is assumed to be the average of the individual unpooled samples. Then, commonly, pooled data consist of a very limited number of observations. Commonly, in order to evaluate pooled data, corresponding parametric assumptions are made. It is a practical issue to test for, e.g., exponentiality or normality using pooled data. For example, an efficient inference can be provided using a single pooled observation, say $X$, if it is accepted that $X$ is normally distributed (Portnoy, 2019).

It turns out that the VpV method can be a valuable tool in developing reasonable and robust testing strategies that is exemplified in Section 2 via Monte Carlo experiments.

In general, the VpV is a function of the data and hence it is a random variable, which too has a probability distribution. In order to study the stochastic character of VpV's, we advance the conventional expected p-value (EPV)-based measure of the performance of a test. The stochastic aspect of the p-value has been well studied by Dempster and Schatzoff (1965) who introduced the concept of the expected significance level under the alternative. Sackrowitz and Samuel-Cahn (1999) developed the approach further and renamed it as the EPV. The authors presented the great potential of using EPV's in various aspects of hypothesis testing.

Comparisons of different test procedures, e.g., a Wilcoxon rank-sum test versus Student's t-test, based on their statistical power is oftentimes problematic in terms of deeming one method being the preferred test over a range of scenarios. One reason for this issue to occur is that the comparison between two or more testing procedures is dependent upon the choice of a pre-specified significance level $\alpha$. One test procedure may be more or less powerful than the other one depending on the choice of $\alpha$ (e.g., Vexler and Yu, 2018). Alternatively, one can consider the EPV approach for comparing test procedures. The EPV is related to the integrated power of a test via all possible values of $\alpha \in [0,1]$. The EPV is one minus the expected power of a test, where the expectation is with respect to an $\alpha$ level which is uniformly [0,1] distributed. Thus, the performance of the test procedure can be evaluated globally using the EPV concept. Smaller values of EPV show better test qualities in a universal fashion.

For example, let the random variable $T$ represent a test statistic depending on data $X$. Assume $F_k$ defines the distribution function of $T$ under the hypothesis $H_k, k = 0, 1$, where the subscript $k$ indicates the null ($k = 0$) and alternative ($k = 1$) hypotheses, respectively. Given $F_k$ is continuous we can denote $F_k^{-1}$ to represent the inverse or quantile function of $F_k$, such that, $F_k(F_k^{-1}(\gamma)) = \gamma$, where $0 < \gamma < 1$ and $k = 0, 1$. In this setting, in order to concentrate upon the main issues, we will only focus on tests of the form: the event $T > C$ rejects $H_0$, where $C$ is a prefixed test threshold. When $F_0$ is known, the p-value can be defined as $1 - F_0(T)$. Then the EPV, $E\{1 - F_0(T) | H_1\}$, is

$$\text{EPV} = \Pr(T^0 \geq T^A), \tag{1.1}$$

where independent random variables $T^0$ and $T^A$ are distributed according to $F_0$ and $F_1$, respectively. The value of the 1-EPV can be expressed in the form of the statistical power of a test through

$$EPV = \Pr(T^0 \geq T^A) = \int_{-\infty}^{\infty} \Pr(T^A \leq t) dF_0(t) = \int_{-\infty}^{\infty} \Pr\{F_0(T^A) \leq F_0(t)\} dF_0(t) \tag{1.2}$$

$$= \int_1^0 \Pr\{1 - F_0(T^A) \geq \alpha\} d(1 - \alpha) = \int_0^1 \left[1 - \Pr\{1 - F_0(T^A) \leq \alpha\}\right] d\alpha = 1 - \int_0^1 \Pr(p - \text{value} \leq \alpha | H_1) d\alpha.$$

Vexler et al. (2018) showed a strong association between the EPV concept and the well-known receiver operating characteristic (ROC) curve methodology (e.g., Vexler and Hutson, 2018, Schisterman, et al., 2005). Such a relationship between the EPV and ROC curve comes in handy for assessing and visualizing the properties of various decision-making procedures in the p-value-based context. This approach was successfully applied to construct optimal multiple testing procedures (Vexler et al., 2018). In Section 3.2 of the present article, the EPV/ROC technique is applied to propose a Youden type criterion for defining optimal tests' critical values.

A wide spectrum of theoretical and applied papers has extensively discussed the old school rule: reject $H_0$ if p-value<0.05 (e.g., .Benjamin, et al., 2018; Wasserstein and Lazar, 2016). We advocate the choice of the significance level $\alpha = 0.05$ using Youden's index and evaluating the likelihood ratio and Bayes Factor (BF) type test procedures. To this end, in particular, we show an intrinsic relationship between the BF test statistic and the BF based on test statistics (Proposition 5).

In Section 3.3, following the ROC curve methodology, we consider a partial EPV (pEPV) to evaluate properties of tests including their unbiasedness. We demonstrate that the conventional power characterization of tests is a partial aspect of the presented EPV/ROC technique.

Section 4 is designed to provide several concluding remarks. We refer to the Appendix for technical derivations and proofs.

## 2. Valid p-Values

Suppose, in statistical analysis of data $X$, we wish to test $H_0: X \sim f_0(x;\theta)$ versus $H_1: X \sim f_1(x)$, where $f_0(x;\theta)$ is a density function that is specified depending on some unknown nuisance parameter $\theta$, the alternative density function $f_1 \neq f_0$ can be assumed to be in an unknown form to state the problem in the nonparametric context. For example, one can consider the goodness-of-fit statement $H_0: X_1 \sim f_0(x;\theta) = \theta \exp(-\theta x), \theta > 0$, vs. $H_1: X_1$ is not exponentially distributed, when $X$ consists of $n$ independent and identically distributed (iid) observations $X_i > 0, i = 1, ..., n$.

Let a statistic $T(\theta)$ based on $X$ be developed to test for $H_0$ vs. $H_1$. In this case, $T(\theta)$ can either contain $\theta$ or have a structure without $\theta$. In order to focus on the main issues, we assume the corresponding decision making procedure can be expressed in such a way that large values of a test statistic $T(\theta)$ are evidence against the null hypothesis $H_0$. Redefine $T(\theta)$'s distribution under $H_k, k = 0,1$, by $F_{T(\theta),k}$. In this framework, in general, we cannot use the p-value in the form

$$p(\theta) = 1 - F_{T(\theta),0}(T(\theta)), \tag{2.1}$$

since $\theta$ is unknown. For example, assuming that $X$ contains iid observations $X_i, i = 1, ..., n$, we have

$$p(\theta) = \int ... \int I\{t(x_1, ..., x_n; \theta) \geq T_{obs}(\theta)\} \prod_{i=1}^{n} f_0(x_i; \theta) dx_1 ... dx_n,$$

where $I$ is the indicator function, $t(x_1, ..., x_n; \theta)$ has a form of the test statistic $T(\theta)$ based on data $(x_1, ..., x_n)$ and $T_{obs}(\theta)$ represents a value of $T(\theta)$ computed using underlying data $(X_1, ..., X_n)$.

The conventional definition of the p-value is

$$p_S = \sup_{\theta \in \Theta} p(\theta), \tag{2.2}$$

where $\Theta$ represents the parameter space for $\theta$ (e.g., Lehmann and Romano, 2006). Unfortunately, definition (2.2) is of rather limited usefulness, since the need to compute the $\sup_{\theta \in \Theta}$ has complicated the problem and, moreover, the supremum is oftentimes too large (in several scenarios $p_S = 1$) to provide a suitable criticism of the null hypothesis (e.g., Bayarri and Berger, 2000). In order to overcome this difficulty, Berger and Boos (1994) and Silvapulle (1996) proposed to denote a valid p-value (VpV) restricting the supremum to $\theta$ in a confidence set for $\theta$. In this setting, let $C_\beta$ define a $1 - \beta$ confidence set for the nuisance parameter $\theta$, under $H_0$. Then, it is suggested to state the VpV in the form

$$p_C = \sup_{\theta \in C_\beta} p(\theta) + \beta. \tag{2.3}$$

In this context, the term "*valid p-value*" signifies that a statistic $p-value \in [0,1]$ is valid if

$$Pr_{H_0}(p-value \leq \alpha) \leq \alpha, \text{ for each } \alpha \in [0,1], \quad (2.4)$$

where $Pr_{H_k}$ denotes the probability under $H_k, k=0,1$. This statement can be applied in the common way to define a level-$\alpha$ decision making procedure. We can decide to reject $H_0$ if and only if the corresponding VpV $\leq \alpha$, having the property (2.4). This principle can simplify and standardize different statistical decision-making policies, providing, e.g., easy strategies for combining test procedures, for example, via the classical Bonferroni method.

P-values defined in (2.2) and (2.3) satisfy (2.4), since denoting the true but unknown $\theta$ by $\theta_0$ and assuming $\beta < \alpha$ we obtain

$$Pr_{H_0}(p_S \leq \alpha) \leq Pr_{H_0}\{p(\theta_0) \leq \alpha\} = Pr_{H_0}\{F_{T(\theta_0),0}(T(\theta_0)) \geq 1-\alpha\} = \alpha \quad \text{and}$$

$$Pr_{H_0}(p_C \leq \alpha) = Pr_{H_0}(p_C \leq \alpha, \theta_0 \in C_\beta) + Pr_{H_0}(p_C \leq \alpha, \theta_0 \notin C_\beta)$$

$$\leq Pr_{H_0}\{p(\theta_0) + \beta \leq \alpha, \theta_0 \in C_\beta\} + Pr_{H_0}(\theta_0 \notin C_\beta) \leq Pr_{H_0}\{p(\theta_0) \leq \alpha - \beta, \theta_0 \in C_\beta\} + Pr_{H_0}(\theta_0 \notin C_\beta)$$

$$\leq Pr_{H_0}\{p(\theta_0) \leq \alpha - \beta\} + \beta = \alpha - \beta + \beta = \alpha,$$

where it is used that $\sup_\theta p(\theta) \geq p(\theta_0)$ and $p(\theta_0)$ is Unif[0,1] distributed under $H_0$.

Unfortunately, concepts (2.2) and (2.3) cannot be applied to some testing problems. This is exemplified in the following section. Note that, oftentimes, the approach of the VpV is addressed by the statistical literature in parametric statistical analysis (e.g., Bayarri and Berger, 2000; Berger and Boos, 1994; Silvapulle, 1996). In the following sections, we consider the VpV approach in nonparametric and parametric settings.

## 2.1. Kolmogorov-Smirnov Tests

In this section we focus on the classical KS goodness-of-fit tests. The presented analysis is relatively clear, and has the basic ingredients for more general cases. Consider the scenario when a statistician is called upon to test some hypothesis about the distribution of a population. If the test is concerned with the agreement between the distribution of a set of sample values and a theoretical distribution we call it a "test of goodness-of-fit".

We begin with examining the goodness-of-fit test for exponentiality based on iid data points $X_1,...,X_n$. We wish to investigate compatibility of the model $H_0 : X_1 \sim f_0(x;\theta)$ $= \theta \exp(-\theta x) I(x>0)$, for some $\theta > 0$, vs. the model $H_1 : X_1$ is not $\sim f_0(x;\theta)$, for all $\theta > 0$. In this case, the usual KS statistic has the form

$$D_n(\theta) = \sup_{0<u<\infty} |1 - \exp(-\theta u) - F_n(u; X)|, \quad (2.5)$$

where $D_n(\theta)$ measures the closeness between the $H_0$-distribution function $1-\exp(-\theta u)$ and the sample (empirical) distribution function $F_n(u; X) = n^{-1} \sum_{i=1}^{n} I(X_i \leq u)$.

It is well-known that the distribution function $Pr_{H_0}\{D_n(\theta_0) \leq u / \text{ the true value of } \theta \text{ is } \theta_0\}$ is independent of $\theta_0$. Then, in this case, it is clear that the p-value

$$p(\theta) = 1 - F_{D_n(\theta),0}(D_n(\theta))$$

$$= \int \ldots \int I\left\{ \sup_{0<u<\infty} |1 - \exp(-\theta u) - F_n(u; x_1, \ldots, x_n)| \geq \sup_{0<u<\infty} |1 - \exp(-\theta u) - F_n(u; X_1, \ldots, X_n)| \right\} \prod_{i=1}^{n} f_0(x_i; \theta) dx_1 \ldots dx_n$$

$$= \int \ldots \int I\left\{ KS_n(x_1, \ldots, x_n) \geq \sup_{0<u<\infty} |1 - \exp(-\theta u) - F_n(u; X_1, \ldots, X_n)| \right\} \prod_{i=1}^{n} f_0(x_i; \theta) dx_1 \ldots dx_n$$

$$= 1 - F_{KS_n,0}(D_n(\theta)),$$

where the statistic $KS_n(x_1, \ldots, x_n)$ based on iid random variables $x_1, \ldots, x_n$ is distributed independently of $\theta$'s values under $H_0$ (see, e.g., Wang et al. 2003, for details). Thus, the VpV's by (2.3) and (2.4) can be computed as

$$p_S = 1 - F_{KS_n,0}\left( \inf_{0<\theta<\infty} D_n(\theta) \right) \text{ and } p_C = 1 - F_{KS_n,0}\left( \inf_{0<\theta\in C_\beta} D_n(\theta) \right) + \beta. \quad (2.6)$$

The following propositions show that when we observe only one single data point ($n=1$) the KS approach is not useful in the context of the VpV method.

**Proposition 1.** Assume we observe only $X = X_1 > 0$. Then the statistic $p_S$ is independent of the data and satisfies $p_S = 1 - F_{KS_n,0}(0.5) = 1$.

In order to apply $p_C$ obtained in (2.6), we denote the maximum $H_0$-likelihood ratio confidence interval for $\theta$ in the form

$$C_\beta = \left[ \theta: \hat{\theta} \exp(-\hat{\theta} X_1) \{\theta \exp(-\theta X_1)\}^{-1} < A_\beta \right],$$

where the maximum $H_0$-likelihood estimator $\hat{\theta}$ of $\theta$ is $1/X_1$ and the threshold $A_\beta$ satisfies

$$Pr_{H_0}\{\theta_0 \in C_\beta / \text{ the true value of } \theta \text{ is } \theta_0\} = Pr\{(\theta_0 x)^{-1} \exp(\theta_0 x - 1) < A_\beta / x \sim Exp(\theta_0)\} = 1 - \beta.$$

In this case, we have the following result.

**Proposition 2.** Assume we observe only $X = X_1 > 0$. Then the statistic $p_C$ is independent of the data and satisfies $p_C = 1 - F_{KS_n,0}(0.5) + \beta = 1 + \beta$, for $\beta \in (0, 0.75)$.

The following arguments show that the issue addressed by Propositions 1 and 2 is not trivial. The problem of making statistical inference based on a single observation has been extensively dealt with in the literature (e.g., Portnoy, 2019). This issue can be considered in both the theoretical and the practical aspects. For example, assume that we survey a statistic $X$, which is a function of unobserved variables $\eta_1,...,\eta_N$, and it is anticipated that $X$ has an asymptotic distribution, say, $\Upsilon$, as $N \to \infty$ (relevant examples related to biological markers evaluations can be found in Section 1 and Vexler and Hutson, 2018: Section 2.4.6). In this case, the problem of investigating compatibility of the model $H_0 : X \sim \Upsilon$ with the observed data point $X$, for a fixed $N$, can be in effect. Someone can propose to test for exponentiality, using the procedure of the form: reject the null hypothesis when $X \geq C$, where $C$ is a test threshold. This procedure can be relatively powerful in many scenarios based on different underlying data distributions. However, it turns out that, in this statement of problem, we cannot define the VpV and control the Type I error rate of the decision making mechanism. Note that, in practice, in order to test for the composite hypothesis of exponentiality, the statistical literature suggests to transform observations $X_1,...,X_n$, for example, applying $X_1 n / \sum_{i=1}^{n} X_i,..., X_n n / \sum_{i=1}^{n} X_i$ (e.g., Henze and Meintanis, 2005). It is clear that we cannot use such invariant (with respect to the parameter $\theta$) transformations when $n=1$.

The above analysis can be adapted to treat different KS type procedures. Consider, for example, the problem of testing for normality based on iid data points $X_1,...,X_n$: $H_0 : X_1 \sim N(\theta,1)$, for some $\theta$, vs. $H_1 : X_1$ is not $\sim N(\theta,1)$, for all $\theta$. In this case, the KS statistic is

$$D_n(\theta) = \sup_{-\infty < u < \infty} \left| \int_{-\infty}^{u} \exp\left(-(z-\theta)^2/2\right) dz / (2\pi)^{1/2} - F_n(u) \right| \text{ with } F_n(u) = n^{-1}\sum_{i=1}^{n} I(X_i \leq u), \quad (2.7)$$

and then

$$p_S = 1 - F_{KS_n,0}\left(\inf_{-\infty < \theta < \infty} D_n(\theta)\right) \text{ and } p_C = 1 - F_{KS_n,0}\left(\inf_{\theta \in C_\beta} D_n(\theta)\right) + \beta, \quad (2.8)$$

where $F_{KS_n,0}$ is the $H_0$-distribution function of the statistic

$$\sup_{-\infty < u < \infty} \left| \int_{-\infty}^{u} \exp\left(-(z-\theta)^2/2\right) dz / (2\pi)^{1/2} - n^{-1}\sum_{i=1}^{n} I(x_i \leq u) \right|$$

based on iid random variables $x_1,...,x_n$ from $N(\theta,1)$ and $F_{KS_n,0}$ is independently of $\theta$'s values; $C_\beta$ defines a corresponding $1-\beta$ confidence set for the parameter $\theta$ under $H_0$.

In this framework, Proposition 3 below displays that the classical KS approach cannot provide the VpV's<1 when we observe only one single data point ($n=1$).

**Proposition 3.** Assume we observe only $X = X_1 > 0$. Then the statistics $p_S$ and $p_C$ are independent of the data and satisfy $p_S = 1 - F_{KS_n,0}(0.5) = 1$, $p_C = 1 - F_{KS_n,0}(0.5) + \beta = 1 + \beta$, where $C_\beta = \{\theta: \exp((X_1 - \theta)^2 / 2) < A_\beta\}$ with $A_\beta: Pr_{H_0}\{\theta_0 \in C_\beta / \text{ the true value of } \theta \text{ is } \theta_0\} = 1 - \beta$ is the maximum $H_0$-likelihood ratio confidence interval for $\theta$.

Suppose the problem is that of testing $H_0: X_1,..., X_n \sim N(0, \theta^2)$, for some $\theta > 0$, vs. $H_1: X_1,..., X_n$ are not $N(0, \theta^2)$ distributed, for all $\theta > 0$. In this scenario, the KS statistic is

$$D_n(\theta) = \sup_{-\infty < u < \infty} \left| \int_{-\infty}^{u/\theta} \exp(-z^2/2) dz / (2\pi)^{1/2} - F_n(u) \right|,$$

and (2.6) defines the VpV when $F_{KS_n,0}$ is the $H_0$-distribution function of the above statistic $D_n(\theta)$ based on iid random variables $x_1,..., x_n$ from $N(0, \theta^2)$, $F_{KS_n,0}$ is independent of $\theta$'s values. Then, one can prove the following result.

**Proposition 4.** Assume we observe only $X = X_1 > 0$. Then the statistics $p_S$ and $p_C$ are independent of the data and we have, for all $\beta \in (0,1)$,

$$p_S = 1 - F_{KS_n,0}(0.5) = 1, \quad p_C = 1 - F_{KS_n,0}\left(D_1\left(\int_{-\infty}^{u_0} \exp(-z^2/2) dz / (2\pi)^{1/2}\right)\right) + \beta \geq 1,$$

where the maximum $H_0$-likelihood ratio confidence interval for $\theta$ has the form

$$C_\beta = \{\theta: \theta |X_1|^{-1} \exp(|X_1|^2 (2\theta^2)^{-1} - 0.5) < A_\beta\},$$

the threshold $A_\beta$ satisfies $Pr\{\eta^{-1/2} \exp(\eta/2 - 1/2) > A_\beta\} = \beta$ with $\eta$ that is a random variable from the $\chi_1^2$ distribution, and $0 < u_0 < 1$ is a root of the equation $u^{-1} \exp(u^2/2 - 1/2) = A_\beta$.

### 2.1.1. Monte Carlo Examples

We evaluated the power of the $p_S$ and $p_C$-based tests (reject $H_0$ if $p_k \leq \alpha$, $k = S, C$) for $H_0: X_1,..., X_n \sim N(0, \theta^2)$, for some $\theta > 0$, vs. $H_1: X_1,..., X_n$ are not $N(0, \theta^2)$ distributed, for all $\theta > 0$, at $\alpha = 0.05$, in the Monte Carlo (MC) manner. The VpV, $p_C$, was defined using $\beta = 0.0005$ and the maximum likelihood ratio interval

$$C_\beta = \left\{\theta > 0: \theta^n \left(\sum_{i=1}^n X_i^2 / n\right)^{-n/2} \exp\left((2\theta^2)^{-1} \sum_{i=1}^n X_i^2 - 0.5n\right) < A_\beta\right\}, \text{ where the threshold } A_\beta$$

satisfies $Pr\{(\eta/n)^{-n/2} \exp(\eta/2 - n/2) > A_\beta\} = \beta$ with $\eta$ that is a random variable from the $\chi_n^2$ distribution. We only exemplify several scenarios where the power of the $p_S / p_C$-based tests is compared with that of the Shapiro-Wilk test for normality combined with the one sample *t*-test for $EX_1 = 0$ in the Bonferroni fashion (the notation SWt denotes this composite test). In the considered nonparametric framework, there are not most powerful decision making mechanisms. Our aim is to demonstrate cases when the $p_S / p_C$-based tests outperform the classical powerful SWt procedure. The following scenarios of source distributions were treated: (A) $X_i = \xi_i - 0.03$, $\xi_i \sim Gamma(1,2)$; (B) $X_i = \xi_i - exp(3)$, $\xi_i \sim LogN(5,1)$; (C) $X_i = \xi_i - 0.45$, $\xi_i \sim \chi_3^2$; (D) $X_i = \xi_i - 0.3$, $\xi_i \sim Weibull(1,5)$; (F) $X_i = \xi_i / \eta_i - exp(3)/2$, $\xi_i \sim LogN(5,1)$ $\eta_i \sim N(2,1)$ (this case is similar to (B), but $E(X_i)$ does not exist in statement (F)), $i = 1,...,n$. At each baseline distribution, the MC experiments were replicated 25,000 times to generate underlying data points $(X_1,...,X_n)$. Table 1 presents the computed MC powers.

**Table 1.** *The Monte Carlo powers of the tests.*

| Test:   | $p_S$  | $p_C$  | SWt    | $p_S$  | $p_C$  | SWt    |
|---------|--------|--------|--------|--------|--------|--------|
| Design: |        | (A)    |        |        | (B)    |        |
| n=7     | 0.7987 | 0.8186 | 0.6204 | 0.9276 | 0.9526 | 0.7193 |
| n=8     | 0.9626 | 0.9660 | 0.7922 | 0.9823 | 0.9908 | 0.8629 |
| n=10    | 0.9963 | 0.9995 | 0.9874 | 0.9993 | 0.9999 | 0.9970 |
| Design: |        | (C)    |        |        | (D)    |        |
| n=7     | 0.6926 | 0.7035 | 0.6315 | 0.7923 | 0.8186 | 0.6153 |
| n=8     | 0.8355 | 0.8459 | 0.7977 | 0.9398 | 0.9717 | 0.7914 |
| n=10    | 0.9641 | 0.9758 | 0.9703 | 0.9952 | 0.9992 | 0.9875 |
| Design: |        | (F) n=7 |       |        | (F) n=20 |      |
|         | 0.7676 | 0.7911 | 0.6735 | 0.9947 | 0.9974 | 0.9822 |

We should note that, for relative large sample sizes, we do not suggest to apply the $p_S / p_C$-based tests in many scenarios with different underlying data distributions. For example, when $n = 50$ and $X_i \sim Cauchy(location = 0, scale = 1/2)$, $i = 1,...,n$, the $p_S / p_C$-based tests and SWt showed powers of 0.06, 0.85 and 0.99 respectively.

**2.2. The Normal Two Sample Problem**

In a similar manner to Section 3 of Berger and Boos (1994) and Section 3.2 of Sackrowitz and Samuel-Cahn (1999), we consider a pedagogical example related to the following scenario. Let $X_1,...,X_n$ be iid $N(\mu_1, \sigma^2)$ and $Y_1,...,Y_m$ be iid $N(\mu_2, \sigma^2)$, independent of the *X*'s. We focus on testing $H_0: \mu_1 = \mu_2$ vs. $H_1: \mu_1 > \mu_2$. If $\sigma^2$ were known, then we could use the t-test statistic

$$T(\sigma) = (\bar{X} - \bar{Y}) / \left\{ \sigma (n+m)^{1/2} (nm)^{-1/2} \right\}, \bar{X} = \sum_{i=1}^{n} X_i / n, \bar{Y} = \sum_{i=1}^{m} Y_i / m,$$

computing the p-value $p(\sigma) = 1 - F_{T(\sigma),0}(T(\sigma))$.

It is clear that $F_{T(\sigma),0}(u) = \Phi(u)$, where $\Phi(u) = \int_{-\infty}^{u} \exp(-z^2/2) dz / (2\pi)^{1/2}$. Then, we can define

$$p_S = 1 - \Phi\left(\inf_{\sigma>0} T(\sigma)\right) \text{ and } p_C = 1 - \Phi\left(\inf_{\sigma \in C_\beta} T(\sigma)\right) + \beta,$$

where $C_\beta$ is the maximum $H_0$-likelihood ratio confidence interval for $\sigma$,

$$C_\beta = \left[\sigma: \sigma^N (\hat{\sigma}_N)^{-N} \exp\left\{(2\sigma^2)^{-1} \sum_{i=1}^{N}(Z_i - \bar{Z})^2 - 0.5N\right\} < A_\beta\right],$$

$(Z_1,...,Z_N) = (X_1,...,X_n, Y_1,...,Y_m)$, $\bar{Z} = \sum_{i=1}^{N} Z_i / N$, $(\hat{\sigma}_N)^2 = \sum_{i=1}^{N}(Z_i - \bar{Z})^2 / N$, $N = n+m$; the threshold $A_\beta$ satisfies

$$Pr\left[\left\{N^{-1}\sum_{i=1}^{N}\left(z_i - N^{-1}\sum_{i=1}^{N} z_i\right)^2\right\}^{-N/2} \exp\left\{0.5\sum_{i=1}^{N}\left(z_i - N^{-1}\sum_{i=1}^{N} z_i\right)^2 - 0.5N\right\} > A_\beta\right] = \beta$$

with iid random variables $z_1,...,z_N \sim N(0,1)$. Note that, the statistic $T(\sigma)$ contains the nuisance parameter $\sigma$ and does not include $\mu_1, \mu_2$. Then, we do not consider $C_\beta$ using $\mu = \mu_1 = \mu_2$ instead of $\bar{Z}$, defining, for example, $p_C = \sup_\mu \left\{\sup_{\theta \in C_\beta} p(\theta) + \beta\right\}$.

In the case of $\bar{X} \leq \bar{Y}$, we have $\inf_{\sigma>0} T(\sigma) = -\infty$ and then $p_S = 1$, whereas when $\bar{X} > \bar{Y}$ we have $\inf_{\sigma>0} T(\sigma) = 0$ obtaining $p_S = 0.5$. Therefore, in this example, $p_S \geq 0.5$, although valid, is useless.

The above analysis leads to $p_C = 1 - \Phi\left(T(\sigma_{L_\beta})I(\bar{X} \leq \bar{Y}) + T(\sigma_{U_\beta})I(\bar{X} > \bar{Y})\right) + \beta$, where $\sigma_{L_\beta}$ and $\sigma_{U_\beta}$ are the lower and upper endpoints of $C_\beta$. (Values of $\sigma_{L_\beta}$ and $\sigma_{U_\beta}$ can be accurately calculated numerically.) So we can propose the VpV based test strategy: reject the null hypothesis if and only if $p_C \leq \alpha$ at a desired user-specified significance level $\alpha$.

**2.2.1. Monte Carlo Examples**

As mentioned above, the example considered in Section 2.2 is called as 'Pedagogical', since, in practice, it is very difficult to compete against the well-known two sample Student's t-test. For example, we used 150,000 MC replications of $X_1,..., X_{10} \sim N(0.7,1)$ and $Y_1,...,Y_{15} \sim N(0,1)$, obtaining the MC powers 0.105 and 0.323 of the $p_C(\beta = 0.005)$-based test and Student's t-test, respectively, at $\alpha = 0.05$. Suppose an investigator anticipate $X_1,..., X_n \sim N(\mu_1, \sigma^2)$ and $Y_1,...,Y_m \sim N(\mu_2, \sigma^2)$. However, a real data corresponds to the scenario: $X_1,..., X_{10} \sim N(0, 2^2)$ and $Y_1,...,Y_{15} \sim N(0,1)$. In this case, the experimental study shows the MC Type I error rates 0.008 and 0.076 of the $p_C$-based test and Student's t-test, respectively, at expected $\alpha = 0.05$. Certainly, if it would be known that $var(X_1) \neq var(Y_1)$, Welch's t-test could be suggested to be applied. Defining $X_i = 1 - \xi_i$, $\xi_i \sim Exp(1)$, $i = 1,...,10$ and $Y_1,...,Y_{15} \sim N(0,1)$ in the simulation study, we calculated the MC Type I error rates 0.004, 0.065 and 0.080 of the $p_C$-based test, Student's t-test and Welch's t-test, respectively, at $\alpha = 0.05$. Then, it seems to be reasonable that, when $'E(X) = E(Y)'$ type conservatism is deemed desirable, to implement the $p_C$-based test.

Note that, in the experiments above, we used the maximum $H_0$-likelihood ratio confidence interval for $\sigma$, anticipating good properties of this likelihood based approach, in the parametric setting. One can simplify the $p_C$-based test, considering

$$C_\beta = C' = \left\{ \sigma^2: 0 \leq \sigma^2 \leq \sum_{i=1}^{N}(Z_i - \bar{Z})^2 / \gamma_\beta \right\},$$ where $\gamma_\beta$ is the $100\beta$ percentile of a $\chi^2_{(N-1)}$

distribution with $N-1$ degrees of freedom. In this case, we observed outputs that were similar to those shown above. For example, in the scenarios examined above: (1) $X_i \sim N(0.7,1)$, $Y_j \sim N(0,1)$ and (2) $X_i = 1 - \xi_i$, $\xi_i \sim Exp(1)$, $Y_j \sim N(0,1)$, $1 \leq i \leq 10$, $1 \leq j \leq 15$, the MC powers of the corresponding $p_C$-based test with $C'$ were calculated as 0.114 and 0.005, respectively.

### 2.3. Remarks

(1) In the frequentist perspective, the meaning of the VpV's is straightforward. Indeed, definitions (2.2) and (2.3) can be regarded as conservative. In general, the statistics extensively evaluated in Bayarri and Berger (2000), e.g., $\int p(\theta)\pi(\theta)d\theta$, where $\pi(\theta)$ is a prior distribution for $\theta$, are not VpV's. In Sections 2.1 and 2.2, we demonstrate MC experiments to provide a practical implementation of the VpV's. (2) In contrast to Bayarri and Berger (2000), we consider test statistics that can contain unknown parameters. In this case, the VpV's convert the decision making procedures into useful mechanisms based on the rule: reject $H_0$ if the corresponding VpV $\leq \alpha$.

## 3. Expected p-Values

### 3.1. Expected valid p-values

Consider first the composite null hypotheses stated in Section 2. In general, expected VpV's have forms that can be different from those investigated in Sackrowitz and Samuel-Cahn (1999). The use of (2.2) and (2.3) leads to the EPV's expressions

$$\text{EPV}_S = E\{\sup_{\theta \in \Theta} p(\theta) / H_1\} \text{ and } \text{EPV}_C = E\{\sup_{\theta \in C_\beta} p(\theta) / H_1\} + \beta.$$

Suppose the problem is to evaluate the KS goodness-of-fit tests. Then, we have

$$\mathrm{EPV}_S = 1 - E\left\{F_{KS_n,0}\left(\inf_{\theta\in\Theta} D_n(\theta)\right)/H_1\right\} = Pr\left(T_S^0 \geq T_S^A\right) \text{ and}$$

$$\mathrm{EPV}_C = 1 - E\left\{F_{KS_n,0}\left(\inf_{\theta\in C_\beta} D_n(\theta)\right)/H_1\right\} + \beta = Pr\left(T_C^0 \geq T_C^A\right) + \beta,$$

where random variables $T_S^0$, $T_S^A$, $T_C^0$ and $T_C^A$ are independent, $T_S^0$ and $T_C^0$ are $F_{KS_n,0}$-distributed, $T_S^A$ and $T_C^A$ are distributed as the statistics $\inf_{\theta\in\Theta} D_n(\theta)$ and $\inf_{\theta\in C_\beta} D_n(\theta)$, respectively, under $H_1$; $F_{KS_n,0}$ and $C_\beta$ are defined corresponding to the statements considered in Section 2.1.

In the framework of the normal two sample problem (Section 2.2), we can determine

$$\mathrm{EPV}_C = 1 - E\left\{\Phi\left(T(\sigma_{L_\beta})I(\bar{X}\leq\bar{Y}) + T(\sigma_{U_\beta})I(\bar{X}>\bar{Y})\right)/H_1\right\} + \beta = Pr\left(T_C^0 \geq T_C^A\right) + \beta,$$

where random variables $T_C^0 \sim N(0,1)$ and $T_C^A$ are independent, $T_C^A$ is distributed as the statistic $T(\sigma_{L_\beta})I(\bar{X}\leq\bar{Y}) + T(\sigma_{U_\beta})I(\bar{X}>\bar{Y})$ based on $\{X_1,...,X_n \sim N(\mu_1,\sigma^2), Y_1,...,Y_m \sim N(\mu_2,\sigma^2)\}$ with $\mu_1 > \mu_2$.

By virtue of the property of expectation of a positive random variable, we have $E(p_k/H_1)$ $= \int_0^1 \{1 - Pr(p_k \leq \alpha/H_1)\} d\alpha$, $k = S,C$, and therefore $\mathrm{EPV}_S$ and $\mathrm{EPV}_C$ are associated with the integrated power of the tests. The quantities $\mathrm{EPV}_S$ and $\mathrm{EPV}_C$ represent one minus the expected power of the corresponding tests, where the expectation is with respect to an $\alpha$ level which is uniformly [0,1] distributed.

In a similar manner to computing the conventional test power functions, in order to obtain values of $\mathrm{EPV}_S$ and $\mathrm{EPV}_C$, the alternative hypothesis $H_1$ should be specified.

### 3.1.1. Monte Carlo Examples

In order to exemplify the $\mathrm{EPV}_S$ and $\mathrm{EPV}_C$ concepts, we used the MC setting shown in Section 2.1.1 regarding the experimental evaluations of the $p_S$ and $p_C$-based tests for $H_0: X_i \sim N(0,\theta^2)$, for some $\theta > 0$, $i = 1,...,n$. The designs (A), (B), (C) and (D) considered in Table 1 were performed. Table 2 displays the MC estimated EPV's.

**Table 2.** *The Monte Carlo EPV's of the tests.*

|         | $\mathrm{EPV}_S$ | $\mathrm{EPV}_C$ | $\mathrm{EPV}_S$ | $\mathrm{EPV}_C$ |
|---------|------|------|------|------|
| Design: | (A)  |      | (B)  |      |
| $n=7$   | 0.0456 | 0.0408 | 0.0407 | 0.0354 |
| $n=8$   | 0.0280 | 0.0251 | 0.0243 | 0.0210 |
| $n=10$  | 0.0105 | 0.0095 | 0.0087 | 0.0077 |
| Design: | (C)  |      | (D)  |      |
| $n=7$   | 0.0566 | 0.0503 | 0.0468 | 0.0409 |
| n=8 | 0.0367 | 0.0326 | 0.0287 | 0.0249 |
| n=10 | 0.0157 | 0.0140 | 0.0110 | 0.0096 |

Thus, in the EPV context, the $p_C$-approach is somewhat better than the $p_S$-approach $(\text{EPV}_C < \text{EPV}_S)$ in the studied scenarios. Table 2 shows that the KS policies discriminate alternative (B) from the model $H_0$ better than alternatives (A), (C) and (D), whereas (C) is a 'worse' scenario in this study.

### 3.2. Why does the significance level $\alpha$ be 5%? -A Youden's index based approach.

As introduced in Section 1, the EPV concept can be treated in light of the ROC curve methodology. Youden's Index is often used in conjunction with the ROC curve technique as a summary measure of the ROC curve (e.g., Schisterman, et al., 2005). It both, measures the effectiveness of a diagnostic marker and enables the selection of an optimal threshold value (cutoff point) for the biomarker. Youden's index, say $J$, is related to the point on the ROC curve which is farthest from line of equality (diagonal line). That is to say, assuming, without loss of generality, that $Z_1,\ldots,Z_n$ and $V_1,\ldots,V_m$ are iid observations from diseased and non-diseased populations, respectively, we have $J = \max_{-\infty < c < \infty} \{\Pr(Z_1 \geq c) + \Pr(V_1 \leq c) - 1\}$.

In this context, we consider a scenario in which $n$ data points $X_1,\ldots,X_n$ are distributed according to the joint density function $f(x_1,\ldots,x_n)$, where $x_1,\ldots,x_n$ are arguments of $f$. In general, the observations do not need to represent values of iid random variables. We would like to classify $X_1,\ldots,X_n$ corresponding to hypotheses of the following form: $H_0$: $\{X_i, i=1,\ldots,n\}$ are from a joint density function $f_0$, versus $H_1$: $\{X_i, i=1,\ldots,n\}$ are from a joint density function $f_1$. We then define the likelihood ratio (LR) as $LR_n = f_1(X_1,\ldots,X_n)/f_0(X_1,\ldots,X_n)$. The LR test based decision rule is to reject $H_0$ if and only if $LR_n \geq C$, where $C$ is a pre-specified test threshold that does not depend on the observations.

In this case, we have $EPV = \Pr(T^0 \geq T^A)$, where independent random variables $T^0$ and $T^A$ are distributed as $LR_n$ under $H_0$ and $H_1$, respectively. Then Youden's approach suggests to find values of $C$ that maximize $\Pr(T^0 < C) + \Pr(T^A \geq C) = \int_0^C f_{LR,0}(u)du + 1 - \int_0^C f_{LR,1}(u)du$, where $f_{LR,k}(u)$ defines the density function of the LR test statistic $LR_n$ under the hypothesis $H_k, k=0,1$. Vexler and Hutson (2018) showed the following result.

**(R1)** For all $u > 0$, $f_{LR,1}(u) = uf_{LR,0}(u)$.

This implies $\Pr(T^0 < C) + \Pr(T^A \geq C) = \int_0^C f_{LR,0}(u)du + 1 - \int_0^C u f_{LR,0}(u)du$ and hence

$$d\{\Pr(T^0 < C) + \Pr(T^A \geq C)\}/dC = f_{LR,0}(C) - C f_{LR,0}(C) = 0$$

gives the optimal test threshold $C = 1$.

The practice of considering normally distributed data when optimal properties of statistical procedures are investigated has historically been common in research (e.g., Fisher, 1922). Let $X_1, ..., X_n$ be iid $N(\mu, \sigma^2)$ and $H_0$ state $\mu = 0$ against $H_1: \mu = \delta \neq 0$. We have $LR_n = \exp\left(\delta \sum_{i=1}^n X_i / \sigma^2 - \delta^2 n / (2\sigma^2)\right)$ and then the Type I error rate is $\Pr\left(2\delta^{-1} \sum_{i=1}^n X_i > n \mid H_0\right)$, when $C = 1$. Following an usual method for evaluating tests' efficiencies (e.g., Lazar and Mykland, 1998), we set $\delta = \tau \sigma n^{-1/2}$. It turns out that, for $|\tau| > 3.3$ ($\delta \approx \pm$ three sigma $\times n^{-1/2}$), the Type I error rate is smaller than $\alpha = 0.05$. The "three sigma" component involved in this analysis can be associated with the so-called three-sigma rule of thumb that expresses a conventional heuristic that nearly all values are taken to lie within three standard deviations of the mean. Three-sigma limit is a statistical calculation that refers to data within three standard deviations from a mean. For example, in business applications, three-sigma refers to processes that operate efficiently and produce items of the highest quality. In this context, e.g., one can refer to well established statistical procedures based on Shewhart charts.

One can also observe that $\Pr\left(2\delta^{-1} \sum_{i=1}^n X_i > n \mid H_0\right) \leq 0.05$, when $\delta \approx \sigma$ and $n \geq 11$ ($n \geq 11$ is a reasonable sample size). Designing a statistical study, it is rational to require that cases with $|E(X \mid H_1) - E(X \mid H_0)| \geq \sigma$ could be detected by the test procedure. It turns out that requesting $n \geq 11$ observations for the study, we could provide the optimal decision making procedure and control the Type I error rate to be <5%, even when $|E(X \mid H_1) - E(X \mid H_0)| = \sigma$.

Note, for example, that, if $\delta = \tau \sigma n^{-1/2}$ and $\tau > 0$, $\Pr\left(2\delta^{-1} \sum_{i=1}^n X_i > n \mid H_0\right) =$

$1 - \Pr\left(\sigma^{-1} \sum_{i=1}^n X_i / n^{1/2} < \tau/2 \mid H_0\right) = 1 - \int_{-\infty}^{\tau/2} \exp\left(-\frac{u^2}{2}\right) du / (2\pi)$ which equals approximately to 0.049, when $\tau = 3.3$. In this case, the power $\Pr\left(2\delta^{-1} \sum_{i=1}^n X_i > n \mid H_1\right)$ is

$1 - \int_{-\infty}^{-\tau/2} \exp\left(-u^2/2\right) du / (2\pi) \approx 0.95$. Then the difference "Power – Type I error" is 0.95-0.049 $\approx 0.9$. Assume, e.g., we select a test threshold $C'$ such that $\Pr(LR_n > C' \mid H_0) =$ $\Pr\left(\sigma^{-1}\sum_{i=1}^{n} X_i / n^{1/2} > \tau/2 + \log(C')/\tau \mid H_0\right) = 0.01$ with $\tau = 3.3$, i.e. $C' \simeq 9.4$. Then we have

$\Pr(LR_n > C' \mid H_1) \simeq 0.83$ and the corresponding difference is 0.83-0.01=0.82.

Equation (R1) shown above can be extended in order to deal with a situation in which, under $H_1$, $n$ data points $X_1,...,X_n$ are distributed according to the joint density function $f_1(x_1,...,x_n;\theta)$ with unknown parameter $\theta$. Let $\pi(\theta)$ represent a prior distribution for $\theta$ under the alternative hypothesis, satisfying $\int \pi(\theta)d\theta = 1$. Using the Bayes Factor (BF) methodology (e.g., Vexler and Hutson, 2018), we define the test statistic

$$B_n = \int f_1(X_1,...,X_n;\theta)\pi(\theta)d\theta / f_0(X_1,...,X_n).$$

**Proposition 5.** For all $u > 0$, $\int f_{B,1}(u)\pi(\theta)d\theta = uf_{B,0}(u)$, where $f_{B,k}$ is the density function of the test statistic $B_n$ under the hypothesis $H_k, k = 0,1$.

The interesting fact is that the BF, $\int f_{B,1}(B_n)\pi(\theta)d\theta / f_{B,0}(B_n)$, based on the BF, $B_n$, comes to be the BF, $B_n$, i.e. $\int f_{B,1}(B_n)\pi(\theta)d\theta / f_{B,0}(B_n) = B_n$.

In Youden's manner, we select a value of the test threshold $C$, maximizing

$$\Pr(B_n < C \mid H_0) + \int \Pr(B_n \geq C \mid H_1)\pi(\theta)d\theta = \int_0^C f_{B,0}(u)du + 1 - \int \int_0^C f_{B,1}(u)du\pi(\theta)d\theta$$

$$= \int_0^C f_{B,0}(u)du + 1 - \int_0^C \int f_{B,1}(u)\pi(\theta)d\theta du.$$

Proposition 5 yields $\Pr(B_n < C \mid H_0) + \int \Pr(B_n \geq C \mid H_1)\pi(\theta)d\theta = \int_0^C f_{B,0}(u)du + 1 - \int_0^C uf_{B,0}(u)du$.

Therefore, we obtain $C = 1$ that maximizes $\Pr(B_n < C \mid H_0) + \int \Pr(B_n \geq C \mid H_1)\pi(\theta)d\theta$.

Note that the BF-based decision making procedure can be asymptotically ($n \to \infty$) associated with the corresponding maximum likelihood ratio test (Vexler and Hutson, 2018).

### 3.2.1. Monte Carlo Experiments

In a parallel with the likelihood ratio test based on normally distributed observations evaluated in Section 3.2, we consider experimentally the following nonparametric examples. Assume we would like to test for normality iid observations $X_1,...,X_n$, using the Shapiro Wilk test. Under the alternative hypothesis, we define $X_i = \xi_i + 3.3n^{-0.5}\eta_i$, $\xi_i \sim N(0,1), \eta_i \sim LogN(0,1.3^2), i=1,...,n$. Generating 100,000 samples of $X_1,...,X_n$ with $n = 100$, for $\alpha = 0.3, 0.1, 0.05, 0.01$, we obtained the differences "MC Power – Type I error ($\alpha$)" as 0.6584, 0.8248, 0.8558 and 0.8450, respectively. In a similar manner, we examined the case with $n = 150$, obtaining the differences as 0.6696, 0.8419, 0.8764, 0.8644. Thus, it seems that $\alpha = 0.05$ is a reasonable selection in these cases. (This conclusions was also confirmed for different distributions of $\eta_i$.)

In the Monte Carlo fashion shown above, we examined the Wilcoxon Rank Sum test based on $X_i = \xi_i + 3.3n^{-0.5}$, $\xi_i \sim N(0,1), i = 1,..,100$. In this case, the differences "MC Power – Type I error ($\alpha$)" had the values 0.6850, 0.8295, 0.8437, 0.7127 corresponding to $\alpha = 0.3, 0.1, 0.05, 0.01$. Applying $X_i = \xi_i + 3.3n^{-0.5}\pi/3^{0.5}$, $\xi_i \sim Logistic(0,1)$, $sd(\xi_i) = 3^{-0.5}\pi, i = 1,..,100$, the differences were obtained as 0.6917, 0.8619, 0.8781, 0.7728.

### 3.3. Partial expected p-values

Expression (1.2) of the EPV considers the weight of the significance level $\alpha$ from 0 to 1. It may appear to suffer from the defect of assigning most of its weight to relatively uninteresting values of $\alpha$ not typically used in practice, e.g., $\alpha \geq 0.1$. Alternatively, we can use the concept of the partial area under the summary ROC curve (AUC) from the ROC methodology to focus on significance levels of $\alpha$ in a specific interesting range by considering the partial expected p-value (pEPV):

$$pEPV = 1 - \int_0^{\alpha_U} \Pr(p-value \leq \alpha | H_1) d\alpha = 1 - \int_0^{\alpha_U} \Pr\{1 - F_0(T^A) \leq \alpha\} d\alpha$$

$$= 1 + \int_0^{\alpha_U} \Pr\{F_0(T^A) \geq 1 - \alpha\} d(1-\alpha) = 1 + \int_1^{1-\alpha_U} \Pr\{F_0(T^A) \geq z\} dz = 1 - \int_{1-\alpha_U}^1 \Pr\{F_0(T^A) \geq z\} dz$$

$$= 1 - \int_{F_0^{-1}(1-\alpha_U)}^\infty \Pr\{F_0(T^A) \geq F_0(t)\} dF_0(t) = 1 - \int_{F_0^{-1}(1-\alpha_U)}^\infty \Pr\{T^A \geq t\} dF_0(t)$$

$$= 1 - \Pr\{T^A \geq T^0, T^0 \geq F_0^{-1}(1-\alpha_U)\}$$

at a fixed upper level $\alpha_U \leq 1$. In general, one can define the function $pEPV(\alpha_L, \alpha_U)$ $= 1 - \int_{\alpha_L}^{\alpha_U} \Pr\{p-value \leq u | H_1\} du$ and focus on $d\{-pEPV(0,\alpha)\}/d\alpha$. Then, in this case, $d\{-pEPV(0,\alpha)\}/d\alpha$ implies the power at a significance level of $\alpha$.

An essential property of efficient statistical tests is unbiasedness (Lehmann and Romano, 2006). An unbiased statistical test satisfies the rule $\Pr(\text{reject } H_0 / H_0) \leq \alpha$ and $\Pr(\text{reject } H_0 | H_1) \geq \alpha$. In parallel with this definition, it is natural to consider the inequality

$pEPV(0,\alpha) \leq 1 - \int_0^\alpha \Pr(p-value \leq u | H_0) du = 1 - \alpha^2/2$, since $p-value \sim Uniform[0,1]$ under $H_0$ (i.e., $\Pr\{p-value \leq u | H_0\} = u, u \in [0,1]$) and we assume $H_1 \neq H_0$. In this case, $d\{pEPV(0,\alpha)\}/d\alpha = -\Pr(\text{reject } H_0 \mid H_0 \text{ is not true})$ and $d(1-\alpha^2/2)/d\alpha = -\alpha$. However, it is clear that the requirement $pEPV(0,\alpha) \leq 1-\alpha^2/2$ is weaker than that of $\Pr(\text{p-value} < \alpha \mid H_1) \geq \alpha$. Then the EPV based concept can extend the conventional power characterization of tests.

Indeed if, for all $\alpha > 0$, $\Pr(\text{p-value} < \alpha \mid H_0 \text{ is not true}) \geq \alpha$ then $pEPV(0,\alpha) \leq 1-\alpha^2/2$. Assume we have a test statistic $T$. To analyze a relationship between the condition $pEPV(0,\alpha) \leq 1-\alpha^2/2$ and the power $\Pr_{H_1}(\text{p-value} < \alpha)$, we present the following proposition.

**Proposition 6.** The inequality $pEPV(0,\alpha) \leq 1-\alpha^2/2$ implies

$$\Pr(\text{p-value} < \alpha \mid H_1) \geq 0.5\alpha + 0.5\alpha f_{T,1}(C_\alpha)/f_{T,0}(C_\alpha),$$

where $C_\alpha$ is the level-$\alpha$ critical value, $C_\alpha = F_0^{-1}(1-\alpha)$, of $T$ and $f_{T,1}/f_{T,0}$ is the likelihood ratio.

For example, when $T = LR_n$, Propositions 5 and 6 provide $\Pr(\text{p-value} < \alpha \mid H_1) \geq 0.5\alpha + 0.5\alpha C_\alpha$. Taking into account the results shown in Section 3.2, it is reasonable to set $\alpha : C_\alpha = 1$. In this case, we conclude $\Pr(\text{p-value} < \alpha \mid H_1) \geq \alpha$.

In the scenario, where $X_1, \ldots, X_n$ are distributed according to $f_1(x_1, \ldots, x_n; \theta)$, under $H_1$, we define the partial expected p-value as $pEPV_\pi(\alpha_L, \alpha_U) = 1 - \int \int_{\alpha_L}^{\alpha_U} \Pr\{p\text{-}value \leq u \mid H_1\} du\, \pi(\theta) d\theta$. In a similar manner to Proposition 6, we have that $pEPV_\pi(0,\alpha) \leq 1-\alpha^2/2$ implies the inequality $\int \Pr(\text{p-value} < \alpha \mid H_1)\pi(\theta)d\theta \geq 0.5\alpha + 0.5\alpha \int f_{T,1}(C_\alpha)\pi(\theta)d\theta / f_{T,0}(C_\alpha)$. Using the BF, $T = B_n$, by virtue of Proposition 5, we obtain $\int \Pr(\text{p-value} < \alpha \mid H_1)\pi(\theta)d\theta \geq 0.5\alpha + 0.5\alpha C_\alpha$ that is $\int \Pr(\text{p-value} < \alpha \mid H_1)\pi(\theta)d\theta \geq \alpha$, when $\alpha : C_\alpha = 1$.

## 4. Concluding Remarks

In this article we have focused on the principle that proper uses of p-values are subject to what investigators could expect from these statistics. Toward this end, the valid statements of p-values and their stochastic aspect have been treated. We have considered the VpV concept in the testing scenarios when composite null models are stated. In this context, we have evaluated the KS goodness-of-fit tests and the normal two sample problem.

We have examined the problem of the goodness-of-fit testability based on a single observation. It turns out that the KS approach is not helpful for obtaining goodness-of-fit level-$\alpha$ tests based on one data point, in many situations. In general, the problem can be formulated as follows: if someone has $k$ observations, can then these data points be tested for being from an assumed distribution function with $h \leq k$ parameters? Further studies are needed to evaluate this framework.

In order to briefly illustrate a practical implementation of the VpV methods, we have exemplified constructions of new test procedures. Although the VpV based tests are conservative, they can be suggested for practical use when underlying data are relatively small.

Attending to the VpV framework, we have advanced the conventional EPV measure of the performance of a test. The expected VpV is shown to be one minus the expected power of a test. We have proposed a Youden's index based principle to define critical values of decision making procedures. In these terms, the significance level $\alpha = 0.05$ can be suggested, in many decision making scenarios. In light of an ROC curve analysis, we introduce partial EPV's to characterize properties of tests including their unbiasedness.

The present article has displayed a small portion of research in the VpV's and EPV's fields. We hope to rekindle a research interest in valid constructions of p-values and evaluations of the stochastic behavior and properties of p-values related to parametric and nonparametric procedures.

## Acknowledgments

Dr. Vexler's effort was supported by the National Institutes of Health (NIH) grant 1G13LM012241-01. I thank Professor Berger for many useful comments and discussions. The author is grateful to the Editor, the Associate Editor and the referees for suggestions that led to a substantial improvement in this paper.

## Appendix

**Proof of Proposition 1.** Consider, for $u > 0$,

$$D(u) = |1 - exp(-\theta u) - I(X_1 \leq u)| = |1 - exp(-\theta u)| I(X_1 > u) + |-exp(-\theta u)| I(X_1 \leq u)$$

$$= \{1 - exp(-\theta u)\} I(X_1 > u) + exp(-\theta u) I(X_1 \leq u),$$

where the function $1 - exp(-\theta u)$ increases and the function $exp(-\theta u)$ decreases with respect to $u > 0$. Then the function $D(u)$ increases, for $u < X_1$, and decreases, for $u \geq X_1$. Thus,

$$D_1(\theta) = sup_{0 < u < \infty} D(u) = max\{1 - exp(-\theta X_1), exp(-\theta X_1)\}.$$

Assume $1 - exp(-\theta X_1) < exp(-\theta X_1)$. In this case, $\theta < log(2)/X_1$ and $D_1(\theta) = exp(-\theta X_1)$ that is a decreasing function with respect to $\theta$. Assume $1 - exp(-\theta X_1) \geq exp(-\theta X_1)$. In this case $\theta \geq log(2)/X_1$ and $D_1(\theta) = 1 - exp(-\theta X_1)$ that is an increasing function with respect to $\theta$. Thus, $D_1(\theta)$ decreases, for $\theta < log(2)/X_1$, and increases, for $\theta \geq log(2)/X_1$. That is, we conclude that $inf\, D_1(\theta) = D_1(log(2)/X_1) = 0.5$. By virtue of (2.6), the proof is complete.

**Proof of Proposition 2.** Define the notation $H_0(\theta_0)$ to indicate the hypothesis $H_0$ when the true value of $\theta$ is $\theta_0$. Now, we will obtain bounds related to the interval $C_\beta$. The function $u^{-1} exp(u-1), u > 0$, has a global minimum at $u = 1$. Then, the threshold $A_\beta$ satisfies $A_\beta > 1$, in order to provide a solution of $Pr_{H_0(\theta_0)}\{(\theta_0 X_1)^{-1} exp(\theta_0 X_1 - 1) < A_\beta\} = 1 - \beta$. Let $0 < u_0 < 1 < u_1$ be roots of the equation $u^{-1} exp(u-1) = A_\beta$. The roots $0 < u_0 < 1 < u_1$ exist, since $A_\beta > 1$ and the function $u^{-1} exp(u-1)$ monotonically decreases, for $0 < u \leq 1$, and increases, for $u > 1$. This behavior of the function $u^{-1} exp(u-1)$ can be used to show that

$$\beta = Pr_{H_0(\theta_0)}\{(\theta_0 X_1)^{-1} exp(\theta_0 X_1 - 1) > A_\beta\}$$

$$= Pr_{H_0(\theta_0)}\{(\theta_0 X_1)^{-1} exp(\theta_0 X_1 - 1) > A_\beta, \theta_0 X_1 \leq 1\} + Pr_{H_0(\theta_0)}\{(\theta_0 X_1)^{-1} exp(\theta_0 X_1 - 1) > A_\beta, \theta_0 X_1 > 1\}$$

$$= Pr_{H_0(\theta_0)}\{\theta_0 X_1 < u_0, \theta_0 X_1 \leq 1\} + Pr_{H_0(\theta_0)}\{\theta_0 X_1 > u_1, \theta_0 X_1 > 1\}$$

$$= Pr_{H_0(\theta_0)}\{\theta_0 X_1 < u_0\} + Pr_{H_0(\theta_0)}\{\theta_0 X_1 > u_1\} = F_{X_1 \sim \theta_0 exp(-\theta_0 x), 0}(u_0 / \theta_0) + 1 - F_{X_1 \sim \theta_0 exp(-\theta_0 x), 0}(u_1 / \theta_0)$$

$$= 1 - exp(-u_0) + exp(-u_1).$$

This defines the system of equations

$$1 - exp(-u_0) + exp(-u_1) = \beta \text{ and } (u_0)^{-1} exp(-u_0) = (u_1)^{-1} exp(-u_1). \quad (A.1)$$

Then, given $\beta$, one can derive values of $u_0$ and $u_1$ that do not depend on values of $\theta$ and provide $Pr_{H_0(\theta_0)}\{u_0 < \theta_0 X_1 < u_1\} = 1 - \beta$. Figure A1 presents numerical solutions of (A.1), depending on $\beta \in (0,1)$. Then, we have $log(2) \in (u_0, u_1)$, for $\beta \leq 0.75$. According to the proof of Proposition 1, $inf_{0<\theta<\infty} D_1(\theta) = D_1(log(2)/X_1) = 0.5$. That is, $inf_{\theta \in C_\beta} D_1(\theta) = inf_{u_0 < \theta X_1 < u_1} D_1(\theta)$
$= D_1(log(2)/X_1) = 0.5$, for $\beta \leq 0.75$. By virtue of (2.6), this completes the proof.

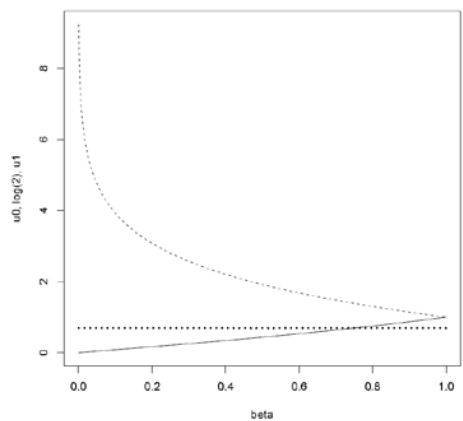

**Figure A1.** The values of $u_0$ (curve '———') and $u_1$ (curve '- - - -'), which satisfy (A.1), plotted against $\beta \in (0,1)$. The line '········' shows $log(2)$.

**Proof of Proposition 3.** It is clear that

$$D_1(\theta) = max\{F_{X_1,0}(X_1), 1 - F_{X_1,0}(X_1)\}, \quad F_{X_1,0}(u) = \int_{-\infty}^{u} exp(-(z-\theta)^2/2)dz/(2\pi)^{1/2}.$$

Assume $F_{X_1,0}(X_1) \geq 1 - F_{X_1,0}(X_1)$, i.e. $F_{X_1,0}(X_1) \geq 1/2$. In this case, $F_{X_1,0}(X_1)$ $= (2\pi)^{-1/2} \int_{-\infty}^{X_1-\theta} exp(-z^2/2)dz$, where $\theta \leq X_1$, and then $D_1(\theta) = F_{X_1,0}(X_1)$ is a decreasing function with respect to $\theta$. Assume $F_{X_1,0}(X_1) < 1 - F_{X_1,0}(X_1)$, i.e. $F_{X_1,0}(X_1) < 1/2$. In this case, $\theta > X_1$ and $D_1(\theta) = 1 - F_{X_1,0}(X_1)$ increases with respect to $\theta$. Thus, $\inf_{-\infty < \theta < \infty} D_1(\theta) = D_1(X_1) = 0.5$.

The point $\theta = X_1$ belongs to the interval $C_\beta = \{\theta: exp((X_1 - \theta)^2/2) < A_\beta\}$. Therefore $\inf_{\theta \in C_\beta} D_1(\theta) = D_1(X_1) = 0.5$. The proof is complete.

**Proof of Proposition 4.** We have

$$D_1(\theta) = max\{F_{X_1,0}(X_1), 1 - F_{X_1,0}(X_1)\}, \quad F_{X_1,0}(u) = \int_{-\infty}^{u/\theta} exp(-z^2/2)dz/(2\pi)^{1/2}.$$

If $X_1 > 0$ then $F_{X_1,0}(X_1) > 1/2$ and $F_{X_1,0}(X_1) > 1 - F_{X_1,0}(X_1)$. In this case, since $D_1(\theta) = F_{X_1,0}(X_1)$ is a decreasing function with respect to $\theta > 0$, $\inf_{0 < \theta < \infty} D_1(\theta) = D_1(\infty) = 0.5$.

If $X_1 < 0$ then $D_1(\theta) = 1 - F_{X_1,0}(X_1)$ is a decreasing function with respect to $\theta > 0$, and $\inf_{0 < \theta < \infty} D_1(\theta) = D_1(\infty) = 0.5$.

Now, we consider $p_C$. Note that, since the function $u^{-1/2} exp(u/2 - 1/2), u > 0$, has a global minimum at $u = 1$, in order to provide a solution of $Pr\{\eta^{-1/2} exp(\eta/2 - 1/2) > A_\beta\} = \beta$, where $\eta \sim \chi_1^2$, the threshold $A_\beta$ should satisfy $A_\beta > 1$. Thus, we have $0 < u_0 < 1 < u_1$ that are roots of the equation $u^{-1} exp(u^2/2 - 1/2) = A_\beta$ and

$$C_\beta = \{\theta > 0: \theta |X_1|^{-1} exp(|X_1|^2/(2\theta^2) - 0.5) < A_\beta\} = \{\theta > 0: u_0 < |X_1|/\theta < u_1\}.$$

According to the above proof scheme, $D_1(\theta)$ is a decreasing function with respect to $\theta > 0$ and then we obtain $p_C = 1 - F_{KS_n,0}(D_1(|X_1|/u_0)) + \beta$, for $\theta \in C_\beta$, where

$$D_1(|X_1|/u_0) = \int_{-\infty}^{u_0} \exp(-z^2/2) dz / (2\pi)^{1/2} I(X_1 \geq 0)$$

$$+ \left\{ 1 - \int_{-\infty}^{-u_0} \exp(-z^2/2) dz / (2\pi)^{1/2} \right\} I(X_1 < 0),$$

since $D_1(\theta) = max\{F_{X_1,0}(X_1), 1 - F_{X_1,0}(X_1)\}$. The distribution function

$\int_{-\infty}^{u} \exp(-z^2/2) dz / (2\pi)^{1/2} = 1 - \int_{-\infty}^{-u} \exp(-z^2/2) dz / (2\pi)^{1/2}$ is symmetric. This implies

$$D_1(|X_1|/u_0) = \int_{-\infty}^{u_0} \exp(-z^2/2) dz / (2\pi)^{1/2}.$$

Now, one can easily use a simple R Code (R Development Core Team, 2002) to compute the accurate Monte Carlo approximations to $p_C = 1 - F_{KS_n,0}\left( D_1\left( \int_{-\infty}^{u_0} \exp(-z^2/2) dz / (2\pi)^{1/2} \right) \right) + \beta$, showing that $p_C \geq 1$ increases when $\beta$ increases. The proof is complete.

**Proof of Proposition 5.** Consider, for nonrandom variables $u$ and $s$, the probability

$$\int \Pr\{u - s \leq B_n \leq u \mid H_1\} \pi(\theta) d\theta = \int E\left[ I\{u - s \leq B_n \leq u\} \mid H_1 \right] \pi(\theta) d\theta$$

$$= \int \left[ \int I\{u - s \leq B_n \leq u\} f_1(x_1, ..., x_n; \theta) dx_1 ... dx_n \right] \pi(\theta) d\theta$$

$$= \int I\{u - s \leq B_n \leq u\} \left[ \int f_1(x_1, ..., x_n; \theta) \pi(\theta) d\theta \right] dx_1 ... dx_n$$

$$= \int I\{u - s \leq B_n \leq u\} \left[ \int \frac{f_1(x_1, ..., x_n; \theta)}{f_0(x_1, ..., x_n)} \pi(\theta) d\theta \right] f_0(x_1, ..., x_n) dx_1 ... dx_n = \int I\{u - s \leq B_n \leq u\} (B_n) f_0.$$

This implies the inequalities

$$\int \Pr\{u - s \leq B_n \leq u \mid H_1\} \pi(\theta) d\theta \leq \int I\{u - s \leq B_n \leq u\} (u) f_0 = u \Pr\{u - s \leq B_n \leq u \mid H_0\} \text{ and}$$

$$\int \Pr\{u - s \leq B_n \leq u \mid H_1\} \pi(\theta) d\theta \geq \int I\{u - s \leq B_n \leq u\} (u - s) f_0 = (u - s) \Pr\{u - s \leq B_n \leq u \mid H_0\}.$$

Dividing these inequalities by $s$ and employing $s \to 0$, we obtain Proposition 5.

**Proof of Proposition 6.** Define the power function $g(u) = \Pr(\text{p-value} < u \mid H_1)$. We have

$\int_0^\alpha g(u) du \geq \alpha^2 / 2$, where $\int_0^\alpha g(u) du = g(u) u \Big|_{u=0}^{u=\alpha} - \int_0^\alpha u w(u) du$, $w(u) = dg(u)/du$. Since

$g(u) = \Pr\{1 - F_{T,0}(T) < u \mid H_1\} = 1 - \Pr\{T < F_{T,0}^{-1}(1-u) \mid H_1\} = 1 - F_{T,1}\left(F_{T,0}^{-1}(1-u)\right)$, we obtain

$w(u) = f_{T,1}(C_u) / f_{T,0}(C_u)$ with $C_u = F_{T,0}^{-1}(1-u)$. It is clear that when $u \nearrow$, the corresponding critical values $C_u \searrow$ and then the likelihood ratio $f_{T,1}(C_u) / f_{T,0}(C_u) \searrow$. This implies

$\alpha^2 / 2 \leq \int_0^\alpha g(u) du = g(\alpha) \alpha - \int_0^\alpha u w(u) du \leq g(\alpha) \alpha - w(\alpha) \int_0^\alpha u du$ that completes the proof.

# References

Bayarri, M. J., and Berger, J. O. (2000) P Values for Composite Null Models, *Journal of the American Statistical Association*, 95, 1127-1142.

Benjamin, D. J., Berger, J. O., Johannesson, M., Nosek, B. A., Wagenmakers, E.-J., Berk, R., et al. (2018) Redefine statistical significance, *Nat. Hum. Behav*. 2, 6–10. doi:10.1038/s41562-017-0189-z

Berger, R. L., and Boos D. D. (1994) P Values Maximized Over a Confidence Set for the Nuance Parameter, *Journal of the American Statistical Association*, 89, 1012-1016.

Dempster, A. P., and Schatzoff, M. (1965) Expected significance level as a sensitivity index for test statistics, *Journal of the American Statistical Association*, 60, 420-436.

Fisher, R.A. (1922) On the mathematical foundations of theoretical statistics, *Philosophical Transactions of the Royal Society A,* 222, 309–368.

Henze, N., and Meintanis, S. G. (2005) Recent and classical tests for exponentiality: a partial review with comparisons, *Metrika*, 61, 29–45.

Ionides, E. L., Giessing, A., Ritov, Y., and Page, S. E. (2017) Response to the ASA's Statement on p-Values: Context, Process, and Purpose, *The American Statistician*, 71:1, 88-89, DOI: 10.1080/00031305.2016.1234977

Lazar, N. A., and Mykland, P. A. (1998) An evaluation of the power and conditionality properties of empirical likelihood, *Biometrika*, 85(3), 523-534.

Lehmann, E. L., and Romano, J. P. (2006) *Testing Statistical Hypotheses*, Springer-Verlag, New York.

Portnoy, S. (2019) Invariance, Optimality and a 1-Observation Confidence Interval for a Normal Mean, *The American Statistician,* 73, 10-15. DOI: 10.1080/00031305.2017.1360796

R Development Core Team. (2002) *R: A language and Environment for Statistical Computing. R Foundation for Statistical Computing*, Vienna, Austria, 2002. http://www.R-project.org.

Sackrowitz, H., and Samuel-Cahn E. (1999) P values as random variables-expected p values, *The American Statistician*, 53, 326-331.

Schisterman, E. F., Perkins, N. J., Liu, A., and Bondell, H. (2005) Optimal cut-point and its corresponding Youden Index to discriminate individuals using pooled blood samples, *Epidemiology*, 16, 73–81.

Schisterman E. F, Vexler A. (2008) To pool or not to pool, from whether to when: applications of pooling to biospecimens subject to a limit of detection. *Paediatric and Perinatal Epidemiology*, 22, 486–496.
23

Schisterman, E. F., Vexler, A., Ye, A. and Perkins, N. J. (2011) A combined efficient design for biomarker data subject to a limit of detection due to measuring instrument sensitivity, *The Annals of Applied Statistics,* 5, 2651-2667.

Silvapulle, M. L. (1996) A Test in the Presence of Nuisance Parameters, *Journal of the American Statistical Association*, 91, 1690-1693.

Vexler, A., and Hutson, A. D. (2018) *Statistics in the Health Sciences: Theory, Applications, and Computing*. CRC Press, New York.

Vexler, A., Schisterman, E. F., and Liu, A. (2008) Estimation of ROC based on stably distributed biomarkers subject to measurement error and pooling mixtures, *Statistics in Medicine,* 27, 280-296

Vexler, A., and Yu, J. (2018) To t-Test or not To t-test? A p-Values-based Point of View in the Receiver Operating Characteristic Curve Framework, *Journal of Computational Biology,* 25, 541–550, DOI: 10.1089/cmb.2017.0216

Vexler, A., Yu, J, Zhao, Y. Hutson, A. D., and Gurevich, G. (2018) Expected P-values in Light of an ROC Curve Analysis Applied to Optimal Multiple Testing Procedures, *Statistical Methods in Medical Research*. In Press. DOI: 10.1177/0962280217704451

Wang, J., Tsang, W. W., and Marsaglia, G. (2003) Evaluating Kolmogorov's Distribution. *Journal of Statistical Software*, 8(18), 1-4. URL http://www.jstatsoft.org/v08/i18/.

Wasserstein, R. L., and Lazar, N. (2016) The ASA's statement on p-values: context, process, and purpose, *The American Statistician,* 70, 129-133.